\documentclass[journal]{IEEEtran}
\ifCLASSINFOpdf
  \usepackage[pdftex]{graphicx}
  \graphicspath{figures}
  \DeclareGraphicsExtensions{.pdf,.jpeg,.png}
\else
\fi
\begin{document}
\bstctlcite{IEEEexample:BSTcontrol}
\title{GHz Rate Neuromorphic Photonic Spiking Neural Network with a Single Vertical-Cavity Surface-Emitting Laser (VCSEL)}

\author{Dafydd Owen-Newns,~\IEEEmembership{Student Member,~IEEE,}
        Joshua Robertson,~\IEEEmembership{Student Member,~IEEE,}
        Mat\v{e}j Hejda,~\IEEEmembership{Student Member,~IEEE,} 
        Antonio Hurtado,~\IEEEmembership{Member,~IEEE}
        
\thanks{The authors are with the Institute of Photonics, Dept. Physics, University of Strathclyde, Technology and Innovation Centre, 99 George Street, G1 1RD, Glasgow (Scotland, UK) (antonio.hurtado@strath.ac.uk).}
\thanks{Manuscript received June 15, 2022; revised X X, X.}%
\thanks{All data underpinning this publication are openly available from the University of Strathclyde KnowledgeBase at https://DOIxxxxxxxxxxxxxxx}%
\thanks{The authors acknowledge this work was supported by the UKRI Turing AI Acceleration Fellowships Programme (EP$/$V025198$/$1), by the European Commission (Grant 828841-ChipAI-H2020-FETOPEN-2018-2020), and by the UK EPSRC (EP$/$N509760$/$1, EP$/$P006973$/$1).}}%

\markboth{Journal of \LaTeX\ Class Files,~Vol.~X, No.~X, June~2022}%
{Shell \MakeLowercase{\textit{et al.}}: Bare Demo of IEEEtran.cls for IEEE Journals}

\maketitle

\begin{abstract}
Vertical-Cavity Surface-Emitting Lasers (VCSELs) are highly promising devices for the construction of neuromorphic photonic information processing systems, due to their numerous desirable properties such as low power consumption, high modulation speed, compactness, and ease of manufacturing. Of particular interest is the ability of VCSELs to exhibit neural-like spiking responses, much like biological neurons, but at ultrafast sub-nanosecond rates; thus offering great prospects for high-speed light-enabled neuromorphic (spike-based) processors. Recent works have shown the use the spiking dynamics in VCSELs for pattern recognition and image processing problems such as image data encoding and edge-feature detection. Additionally, VCSELs have also been used recently as nonlinear elements in photonic reservoir computing (RC) implementations, yielding excellent state of the art operation. This work introduces and experimentally demonstrates for the first time the new concept of a Ghz-rate photonic spiking neural network (SNN) built with a single VCSEL neuron. The reported system effectively implements a photonic VCSEL-based spiking reservoir computer, and demonstrates its successful application to a complex nonlinear classification task. Importantly, the proposed system benefits from a highly hardware-friendly, inexpensive realization (built with a single VCSEL and off-the-shelf fibre-optic components), for high-speed (GHz-rate inputs) and low-power (sub-mW optical input power) photonic operation. These results open new pathways towards future neuromorphic photonic spike-based information processing systems based upon VCSELs (or other laser types) for novel ultrafast machine learning and AI hardware.
\end{abstract}

\IEEEpeerreviewmaketitle

\section{Introduction}

\IEEEPARstart{I}{nformation} processing platforms, inspired by biological neural networks, are at the forefront of both academic and industrial research as major efforts are being made to improve the energy efficiency and speed of machine learning and artificial intelligence (AI) hardware. These so-called neuromorphic systems, are being developed as alternative (non-Von Neumann) processing architectures, and make use of artificial neurons and artificial neural networks (ANNs) to perform brain-inspired information processing. ANNs are highly parallel systems that use layers of multiple nodes (neurons) to efficiently realize nonlinear transformations which, once trained, can be used to perform complex processing tasks \cite{Alzubaidi2021}. With the push for higher energy efficiencies and faster operating speeds, the attractive properties of the photonic platform (e.g. large bandwidths, low cross-talk, low power consumption and ultrafast speeds) have drawn increasing interest from the research community searching for novel light enabled neuromorphic processing systems \cite{Shastri2021}. 

Impressively, despite the recent emergence of the field of neuromorphic photonics, multiple reports of photonic ANN accelerators, based on technologies such as micro-ring weight banks \cite{Tait2017,DeLima2019,Mehrabian2018}, modulators \cite{Xu2019}, and phase change materials \cite{Feldmann2019,Feldmann2021}, amongst others, have been produced, realizing systems capable of multiple different processing tasks. Yet, a quintessential photonic device -the semiconductor laser- has seen a significant and increasing amount of research attention, in part thanks to its capability to provide dynamical behaviours suitable for individual artificial neurons, as well as for ANN implementations (see \cite{Prucnal2016} for a review).
More specifically, one type of semiconductor laser, the Vertical Cavity Surface Emitting Laser (VCSEL), has recently demonstrated the capability to operate both as an ultrafast artificial photonic spiking neuron \cite{Robertson2019,Robertson2017,Hejda2020} and as the key-nonlinear element in photonic ANN realizations, through the technique of reservoir computing (RC) \cite{Vatin2020,Bueno2021,Porte2021}.

Importantly, thanks to their compact structures, vertical light emission, low cost and technology maturity, VCSELs are ubiquitously deployed in our society for disparate industrial uses, spanning from automotive sensors, to supermarket bar code scanners and telecommunications, just to name a few. However, their investigation as potential candidates for photonic artificial neurons for brain-inspired computing systems has been building over the past few years. Their unique attributes (e.g. high modulation speeds, low energy operation, reduced costs, compactness, high coupling efficiency to optical fibres, ease of integration in 2-dimensional arrays, etc. \cite{Koyama2006}) make them exciting candidates for fast, low energy neuromorphic hardware elements. Most interestingly, these devices have produced multiple dynamical behaviours analogous to those exhibited by biological neurons in the brain. It has been shown that off-the-shelf VCSELs can be made to elicit fast (100\,ps-long) action potential-like spiking responses at high (GHz) rates \cite{Hurtado2015}. Further, additional reports of neuron-like properties, such as threshold-and-fire operation \cite{Robertson2019}, spike  inhibition \cite{Robertson2017}, spike rate-encoding and refractoriness \cite{Hejda2020}, now place VCSELs as highly promising candidates for the realization of photonic artificial spiking neurons.

Moreover, beyond the promising realization of single artificial photonic neurons, VCSEL-based ANNs, developed through a technique commonly known as reservoir computing (RC), have recently appeared in literature \cite{Vatin2020,Bueno2021,Porte2021}. The RC technique uses the fixed internal connections of non-linear systems (such as VCSELs) to create what is known as a reservoir. This can be treated as an ANN that utilises the intrinsic processing capability of the dynamical system to perform machine learning functionalities. The fixed internal connections of the reservoir mean that only the output weights of the  ANN require calculation; hence significantly improving the speed and efficiency of network training \cite{article}. Within the concept of RC are extreme learning machines (ELMs) \cite{HUANG2006489}. ELMs are purely feed-forward neural networks that operate on the same principles of a fixed reservoir of nodes, and a single output layer that needs training. The simplified architecture and training procedure makes the implementation of RC/ELM systems in photonic hardware highly desirable, and now photonic platforms based on numerous devices \cite{Kuriki2018,Nakajima2021,Sackesyn2021,Bueno2021,Porte2021} (also recently including VCSELs), have successfully performed information processing tasks such as image classification, time series prediction and signal conditioning with state of the art performance. However, to date neuromorphic photonic systems, such as laser-based photonic reservoir computers, have not demonstrated spiking neural network (SNN) architectures using excitable neural-like spikes for information processing.

This work combines for the first time the concept of artificial photonic VCSEL-based spiking neurons with that of photonic RC, to create a radically-new photonic SNN architecture using fast (sub-ns) excitable optical spikes to process information. The proposed architecture, based on the RC/ELM paradigm, delivers a photonic SNN comprised of $>$1000 ultrafast spiking nodes (at 250\,ps/node), whilst also yielding a hardware-friendly platform using a single VCSEL for the (spiking) reservoir. We report the successful application of this VCSEL-based SNN in a complex nonlinear classification task (Fisher's Iris flower classification) achieving $>$97\% accuracy. These results offer exciting prospects for novel VCSEL-based SNNs for future cheap, low-energy, high-speed and hardware-friendly photonic neuromorphic computing platforms.

In this paper, we first provide in Section II an introduction to the recent experimental reports of VCSEL-based neuromorphic processing systems. In Section III we next describe the technique and characteristics of the experimental layout developed to implement the novel VCSEL-based SNN architecture for photonic spike-based RC. Further, in Section III we then describe the complex nonlinear task (classification of Iris flower species) selected to demonstrate the successful performance of the system, and report the effect of altering system parameters (such as the number of network nodes) on the performance of the spike-based RC system. Finally, in Section IV we provide the conclusions of this work.

\section{VCSELs for Neuromorphic Processing Systems}
\subsection{Information Processing with spiking VCSEL neurons}
\label{Sect:A}

\begin{figure}[!t]
\centering
\includegraphics[width=3.2in]{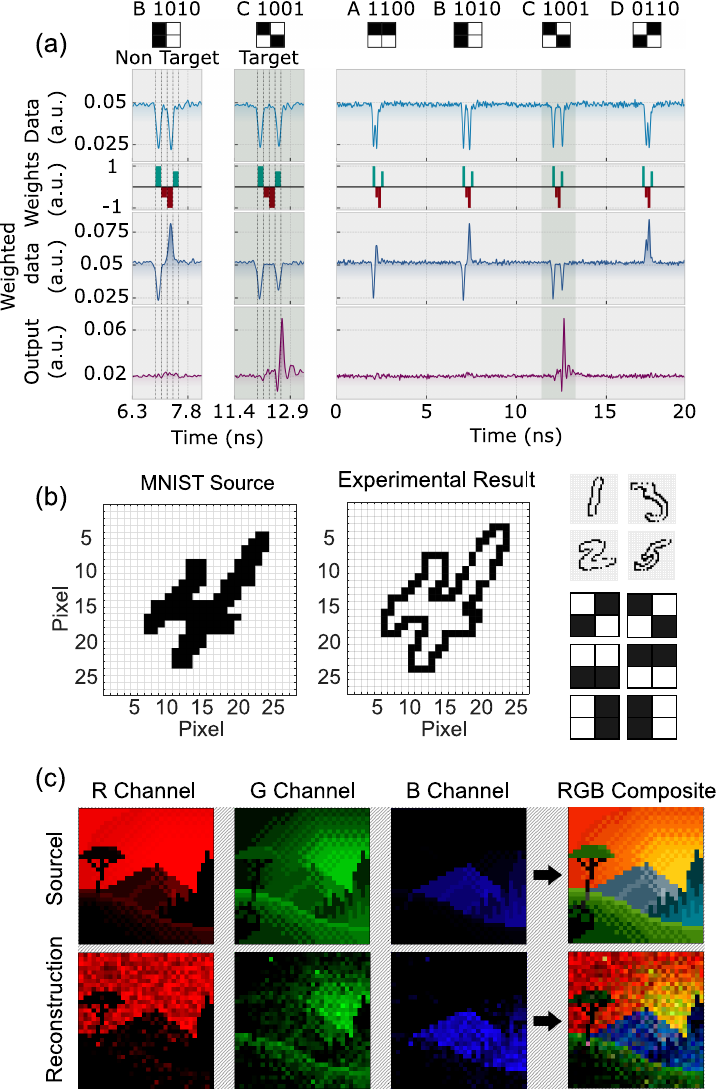}
\caption{Examples of information processing with a spiking VCSEL neuron. a) 4-bit pattern recognition with a spiking VCSEL neuron. Input binary patterns are weighted and optically encoded into the injection of the VCSEL neuron. The VCSEL neuron triggers a fast spiking response for the target pattern (C 1001). b) Image processing with a spiking VCSEL neuron. The integrate-and-fire edge detection of 5000 MNIST handwritten digit images was performed in combination with a SNN reaching a classification accuracy of 96.1\%. c) The spike rate encoding and decoding of image colour channels and the final spike rate reconstruction. a) - c) Reprinted with permission from \cite{Robertson2020Ultrafast,Robertson2022Ultrafast,Hejda2021}, under the terms of the Creative Commons CC BY license.}
\label{fig:Sect2Fig1}
\end{figure}

VCSELs operating at key telecom wavelengths (1300\,nm and 1550\,nm) have in recent years generated reports of neuron-like behaviours (e.g. spike activation and inhibition, refractoriness, threshold-and-fire, integrate-and-fire, etc.) with their potential use as artificial spiking photonic neurons having been outlined \cite{Hurtado2015,Robertson2019,Robertson2017,Hejda2020}. Therefore, these early works have since opened the way for new reports of spiking VCSEL neurons capable of brain-inspired processing tasks (e.g. temporal pattern recognition and image edge feature detection) \cite{Robertson2020Ultrafast,Zhang2021Pyramidal,Robertson2020Image,Zhang2021Binary,Robertson2022Ultrafast,Hejda2021}. The first experimental processing task performed by a spiking VCSEL neuron was the coincidence detection of two input pulses and the subsequent recognition of 4-bit input patterns \cite{Robertson2020Ultrafast}. In this report VCSEL neurons were found to behave analogously to integrate-and-fire neuronal models that combine the influence of multiple inputs arriving within a short temporal window (approx. 1\,ns). It was found that the VCSEL neuron only triggered a fast 100\,ps-long spiking response when a sufficient number of sub-threshold inputs occurred within a short integration time. Using this neural-like behaviour, a VCSEL neuron then demonstrated that coincidence detection (within a short sub-ns temporal window) could be performed with fast optical inputs, realizing all-optical spike-based alarm triggering \cite{Robertson2020Ultrafast}. The recognition of 4-bit input patterns (see Fig.\,\ref{fig:Sect2Fig1}\,(a)) was then demonstrated using the hardware-friendly single spiking VCSEL neuron system. Different 4-bit binary (1 or 0) input patterns at fast 80\,ps/bit rates, were time multiplexed into bursts of input pulses, weighted by a user defined set of weights, and injected into the VCSEL neuron. The $<$\,1\,ns-long input pulse bursts were then integrated by the VCSEL neuron, triggering spikes for the recognition of target patterns only. The system achieved high recognition efficiencies for 14 different patterns of 1, 2 and 3 active bits. These experimental results showcased for the first time that the neuronal behaviours of VCSELs could be harnessed and implemented in order to achieved an experimental optical spike-based pattern recognition task at near GHz rates. Similarly, spiking VCSEL neurons subject to dual modulated optical injection have been used to experimentally achieve all-optical XOR classification functionality, similar to the behaviour of biological pyramidal neurons \cite{Zhang2021Pyramidal}.

More recently, artificial spiking VCSEL neurons have also been implemented to realize image processing tasks, such as image edge-feature detection \cite{Robertson2020Image,Zhang2021Binary}. In this process, kernel operators are used to pre-process digital images before injecting them into the VCSEL neuron, which triggers fast (100\,ps-long) optical spikes for the desired target feature. It was shown that by combining multiple kernel operators the image gradient magnitude could be extracted, subsequently detecting all edge features in a single run of the image at the high rate of 1.5\,ns per pixel \cite{Robertson2020Image}. Further, a binary convolution approach using again a single VCSEL neuron, was shown to calculate image intensity gradients with high processing efficiency and low power requirements, all with a hardware-friendly system \cite{Zhang2021Binary}. Similar single VCSEL neuron systems have also reported the capability to rate-code the pixel colour information of RGB images in the (GHz) firing rates of optical spike trains \cite{Hejda2021}. The demonstration, shown in Fig.\,\ref{fig:Sect2Fig1}\,(c), used the input-dependent spike rate of the VCSEL neuron to assign different colour intensities a spike firing frequency. This created a spike-timing dependent information encoder that converted an RGB image into fast (sub-nanosecond) spike trains. The reconstruction of the image was performed (Fig.\,\ref{fig:Sect2Fig1}\,(c)) to demonstrate the potential for spike-based image encoding and processing. Finally, edge-feature detection and classification of MNIST handwritten digits was recently reported with a single experimental spiking VCSEL neuron working together with a software-implemented SNN \cite{Robertson2022Ultrafast}. In this demonstration a total of 5000 images were processed using the VCSEL neuron and 6 symmetrical kernels, identifying edge features with fast spiking responses (Fig.\,\ref{fig:Sect2Fig1}\,(b)). The generated optical spike trains were then directly fed into the software-implemented SNN which provided a classification of the hand-written digits, achieving an average classification accuracy of 96.1\% \cite{Robertson2022Ultrafast}. These demonstrations therefore highlight a few recent reports of promising spike-based image processing systems built using hardware-friendly neuromorphic VCSELs.

\subsection{Information Processing with VCSEL-based Photonic Reservoir Computers}
\label{Sect:B}

\begin{figure}[!t]
\centering
\includegraphics[width=3.5in]{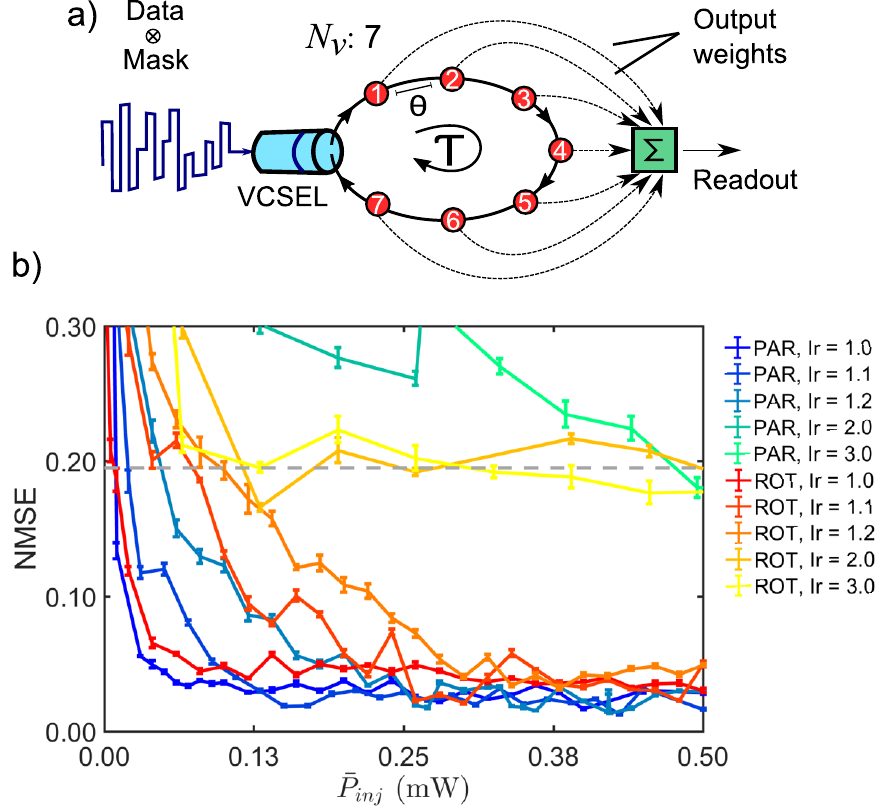}
\caption{Information processing with a VCSEL-based time-delay reservoir (TDR). (a) Schematic of a TDR built using a single VCSEL and a feedback loop. The output of the VCSEL is time-multiplexed to create virtual nodes. Nodes are sampled, weighted and combined to provide the system readout. (b) The performance of a VCSEL-based RC system on a chaotic timeseries prediction task. The normalised mean square error (NMSE) is plotted for parallel and rotated polarization configurations at multiple VCSEL bias currents ($I_{r}$). Overall low error performance with an NMSE = 0.012 was achieved for a parallel-polarization configuration. (b) reproduced © 2022 IEEE. Reprinted, with permission, from \cite{Bueno2021}.}
\label{fig:Sect2Fig3}
\end{figure}

Another promising avenue for VCSEL-based information processing systems, that has recently received pressing research attention, is photonic reservoir computing (RC). Unlike spiking photonic neuromorphic systems, photonic RC systems do not rely on neural-like signals for operation. Instead, laser-based RC systems use the inherent nonlinear dynamical responses of lasers (VCSELs in this case) subject to external optical injection and$/$or feedback to process information. Recently, VCSEL-based RC systems have been demonstrated on two different experimental architectures, namely time delay reservoirs (TDRs) \cite{Vatin2018,Vatin2019,Vatin2020,Bueno2021} and spatial-temporal reservoirs \cite{Porte2021,Skalli2021}. Approaches based on the TDR architecture (shown in Fig.\,\ref{fig:Sect2Fig3}\,(a)), create an interconnected network of virtual nodes using a single VCSEL and an optical delay line (round trip time $\tau$). Input information is first weight-masked and optically injected into the VCSEL, which acts as a non-linear element transforming inputs and feedback signals continuously. The continuous intensity output of the VCSEL is then sampled at set periods ($\theta$) to read the output state of each individual virtual node ($N_v$ total nodes). Finally, a set of output weights are applied to the nodes before each is linearly combined ($\Sigma$) into the final output state of the reservoir. In these TDRs, the nodes of the network are all internally connected given $\theta$ occurs on a timescale smaller than the (VCSEL’s) non-linearity \cite{Appeltant2011}. The $\tau$-long feedback loop in the system works to create memory, providing additional network connections across time \cite{Hulser2022}. It has been shown that these VCSEL-based TDRs not only provide a hardware-friendly solution to photonic RC systems (using only a single non-linear element), but in fact deliver state of the art performance at benchmark classification and time series prediction tasks \cite{Vatin2018,Vatin2019,Vatin2020,Bueno2021}. It was also found that the intrinsic polarization properties of  VCSELs provided additional avenues to enhanced performance via the control of light polarization in both injection and feedback channels. This feature of VCSEL-based TDRs has since been investigated, as shown in Fig.\,\ref{fig:Sect2Fig3}\,(b), revealing that the performance of tasks like the chaotic Mackey-Glass time series prediction, can be significantly improved given the appropriate polarization configuration \cite{Bueno2021}. Furthermore, the polarization properties of these systems has also been exploited to demonstrate parallel processing, completing two non-linear channel equalisation tasks with a single VCSEL \cite{Vatin2020}. VCSEL-based TDRs are therefore extremely powerful and hardware-friendly neuromorphic processing tools.

VCSEL-based spatial-temporal reservoirs implement individual nodes as physical points in space (spatially-multiplexed). In a recent first report \cite{Porte2021}, a spatial-temporal reservoir based on a large area VCSEL (LA-VCSEL) was showcased. Here, the light output from different points on the surface was spatially-multiplexed creating the individual network nodes, which were interconnected via surface carrier interactions and optical diffraction. This LA-VCSEL-based system achieved low error performance during a 3-bit binary header classification task and a further report was generated studying the ideal conditions for reservoir operation \cite{Skalli2021}. Overall, these reports demonstrate that VCSELs offer an exciting platform for photonic RC systems using the inherent nonlinear dynamical behaviours produced in these devices, yielding high performance across diverse complex tasks (non-linear channel equalisation, time series prediction, etc.) in different architectures.

\section{Photonic SNN Architecture For Implementing a Photonic Spiking Reservoir Computer}
\subsection{VCSEL-based SNN Structure and Experimental Setup}
\label{Sect:3A}

\begin{figure}
    \centering
    \includegraphics[width=3.2in]{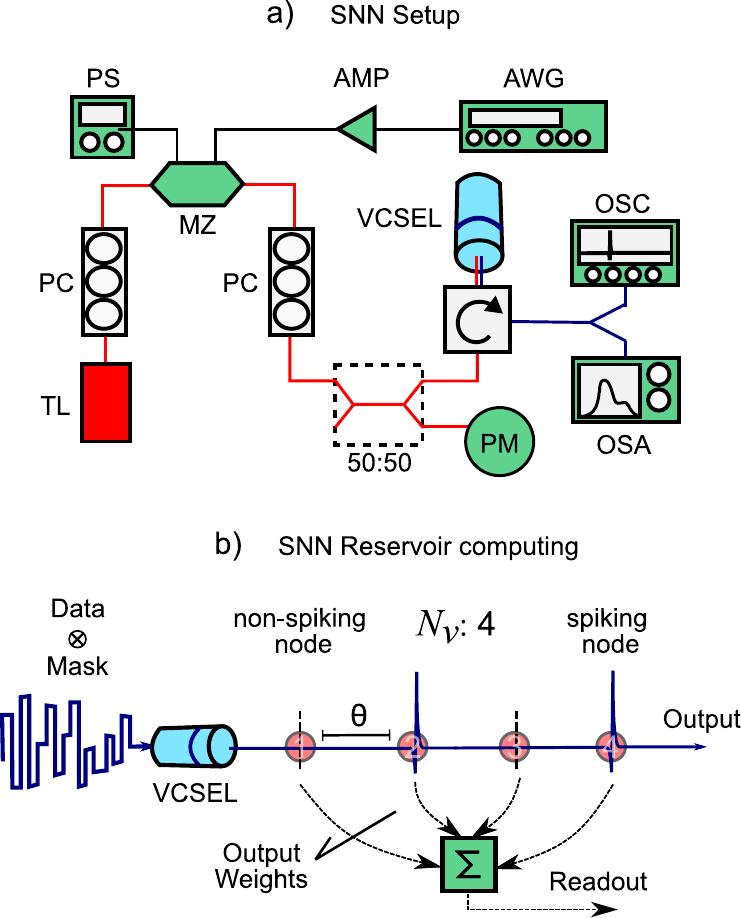}
    \caption{Experimental setup and SNN operation. a) The experimental setup for the spiking reservoir computer/SNN. An external Tuneable Laser (TL) injects intensity-modulated light into the VCSEL via an optical coupler and circulator. Injection light polarization is controlled by polarization controllers (PCs). The masked input signals are generated by an arbitrary waveform generator (AWG) and amplified electrically (AMP) before entering the DC power supply (PS) -controlled Mach-Zehnder intensity modulator (MZ). The optical spiking signals produced by the VCSEL are analysed using a 16\,GHz oscilloscope (OSC) and a optical spectrum analyser (OSA). b) The operation of the spiking RC system. Input data is multiplied by a random mask before injection into the VCSEL. The VCSEL is operated near the injection locking boundary where inputs can produce non-linear spiking responses. Inputs occurring within the refractory period/non-linear timescale of the spiking dynamics ($\sim$1\,ns) are internally connected, hence producing unique spiking patterns at SNN output. The virtual nodes are measured from the optical output of the VCSEL at intervals of $\theta$ ($\theta = 250$\,ps), internally connecting 4 sequential virtual nodes. The output weights are applied to each node and linearly combined to provide the readout of the reservoir.}
    \label{fig:exp-setup}
\end{figure}

As previously outlined, reservoir computers and ELMs are two similar types of ANNs whose structures are recurrent (RC) and feed-forward (ELM). These ANNs each host a hidden layer of interconnected nodes whose weights and parameters do not need to be adjusted, leaving only the single output layer requiring training \cite{article,HUANG2006489}. The structure of ELMs are therefore close to those of reservoir computers, but lack any feedback connections or recurrence. As pointed out in Section \ref{Sect:B}, VCSELs have demonstrated they are ideal candidates for photonic RC (or ELM) systems given their nonlinear dynamical behaviours and inherent advantages (e.g. compactness, high speed, low-energy, reduced costs, etc). In parallel, as reported in Section \ref{Sect:A}, VCSELs can operate as artificial spiking neurons as they exhibit nonlinear neural-like behaviours (e.g. leaky-integrate and fire responses and refractoriness). Remarkably, in the same way it is performed in VCSEL-based TDR systems \cite{Bueno2021}, the neuron-like responses in spiking VCSELs also permits the coupling of multiple time-multiplexed virtual nodes for the creation of complex, interconnected ANN structures. The coupling of virtual nodes in this way (via time-multiplexing) is possible if the temporal duration of the nodes ($\theta$) is set shorter than the time-scale of the neuronal spiking non-linearity (approx. 1\,ns in the investigated VCSELs \cite{Hejda2021}). Importantly, the coupled time-multiplexed nodes now deliver spike firing responses instead of a continuous output, thus using truly neural-like signals for operation. Using this technique (depicted graphically in Fig. \ref{fig:exp-setup}\,(b)), a single VCSEL can deliver a system of coupled spiking neurons (with sub-nanosecond temporal operation) effectively yielding a novel photonic SNN architecture, operating as a spiking photonic RC platform. In Fig. \ref{fig:exp-setup}\,(b) no recurrent/feedback connections are created, thus the system will operate as a photonic spiking ELM which only requires the hardware implementation of a single spiking VCSEL. In this approach, the input signals entering the VCSEL-based SNN (spiking photonic RC system) are chosen (for this proof-of-concept demonstration) to be analogous to those used previously in traditional non-spiking laser-based photonic RC systems (i.e. continuously modulated light input) \cite{Bueno2021}. Yet, we believe operation with other types of input signals (such as pulsed or spiking inputs) should also be possible and will be investigated in the future. Importantly, now (as seen in Fig. \ref{fig:exp-setup}\,(b)), the signals used for computation (at the output of the VCSEL-based reservoir) are event-based (sub-nanosecond long optical spikes) and binary. The value of the system's output nodes are directly determined by the firing (binary ‘1’) or non-firing (binary ‘0’) of a short 150\,ps optical spike within the timeframe $\theta$ of the virtual node.

The experimental setup used to build the VCSEL-based SNN (a spiking photonic RC system) is shown in Fig. \ref{fig:exp-setup}\,(a). This all-optical fibre setup consists of a 1300\,nm tuneable laser (TL) whose output is externally modulated by a Mach-Zehnder (MZ) intensity modulator. The modulated TL light’s is subsequently optically-injected into an off-the-shelf 1300\,nm VCSEL operating as the spiking photonic RC system. The optical injection power is measured using a power meter (PM) and the corresponding output of the VCSEL-based SNN and spiking RC system is measured using an optical spectrum analyser (OSA), a fast ($>$9\,GHz) amplified photodetector and a high-speed (16\,GHz) real-time oscilloscope (OSC). An arbitrary wave generator (AWG) operating at 12\,GSa/s generated the input waveforms to the spiking RC system. The output signals of the AWG are amplified with a 14\,GHz electrical amplifier (AMP) before their injection into the RF input of the MZ modulator, which encodes the input waveforms into the TL’s light.  The DC bias input of the MZ is set using a power supply (PS). The 1300\,nm VCSEL used in the experiments had a measured lasing threshold current of 1.4\,mA at room temperature (293\,K). The VCSEL’s spectrum showed two peaks, corresponding to the two orthogonal polarizations of the fundamental transverse mode of the device (see \cite{Robertson2019} for exemplar device spectral characteristics). The polarization of the TL’s injected light is matched to that of the subsidiary attenuated mode of the VCSEL (using Polarization Controllers, PCs) and injected such that the VCSEL injection-locks to the TL’s light emission. Under these conditions, as described in detail in previous works (see for example \cite{Robertson2019} and references therein) the VCSEL can be made to fire fast optical spikes in response to intensity modulated optical injection. This neural-like behaviour is used here to develop a first spiking photonic RC system, in which the spiking patterns elicited in the VCSEL will depend on the specific input data waveforms.

\subsection{Nonlinear Classification Task Operation}
\label{Sect:3B}

\begin{figure}
    \centering
    \includegraphics[width=3.45in]{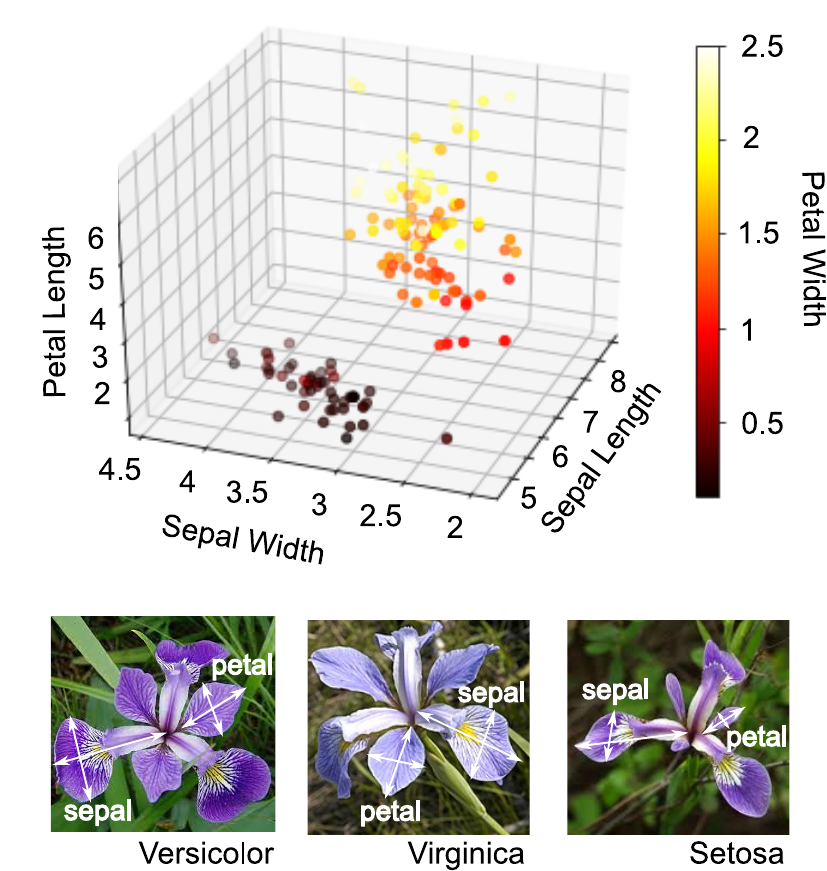}
    \caption{Scatter plot of Fisher's Iris data used in the classification task \cite{Fischer1996}. One class (Setosa) is linearly separable, the other two (Versicolor \& Virginica) are less easily separable.}
    \label{fig:iris-scatter}
\end{figure}

\begin{figure}
    \centering
    \includegraphics[width=3.45in]{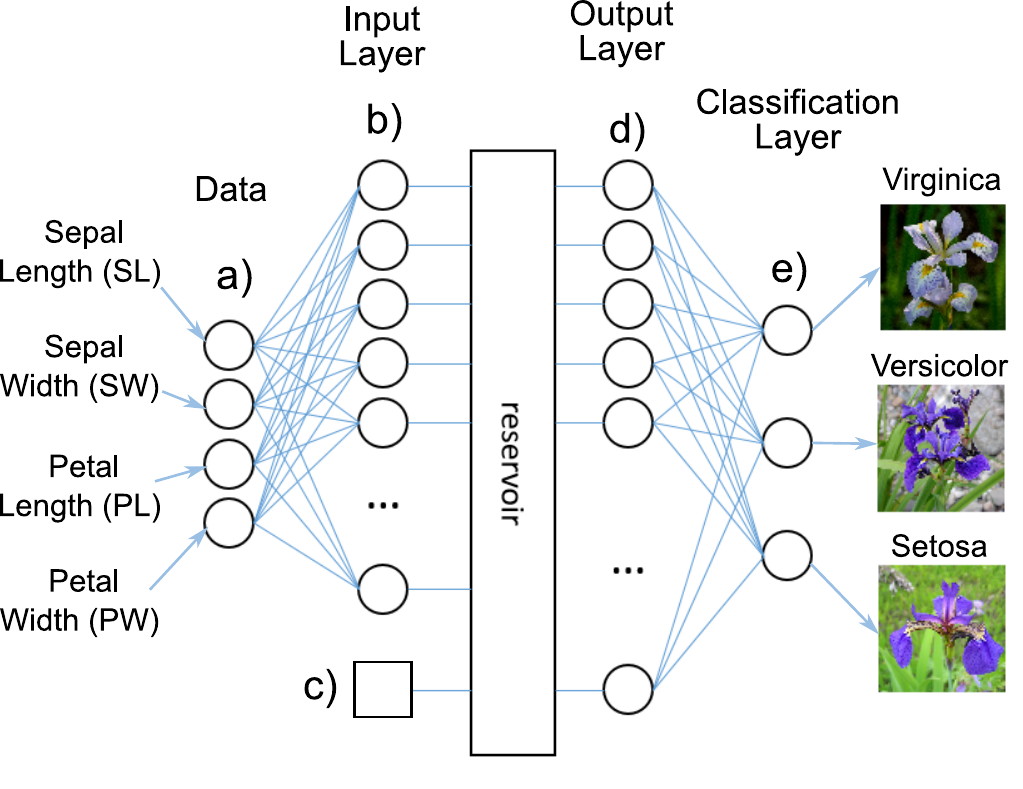}
    \caption{The logical architecture of virtual nodes in the SNN. Each node is distributed in time, separated by $\theta$. The reservoir represents the nonlinear coupling between each of the input nodes. a) The 4 data variables (SL, SW, PL, PW). b) The input layer where values are calculated by applying a random mask to the data. c) A disconnected input representing the time-spacing introduced between data points. d) The output layer where binary values represent whether a spike has or has not occurred. e) The classification layer that provides the probability a data point belongs in each class. The total number of nodes in the both the input and output layer of the SNN is equal to $N_v$. The only weights that require adjusting (training) are the connections between d) and e).}
    \label{fig:reservoir-architecture}
\end{figure}

The task selected to demonstrate the performance of the new photonic VCSEL-based spiking RC system is that of Fisher’s 1936 iris flower dataset, a well-known benchmark Machine Learning classification task \cite{Fisher:1936}. Each data point in this dataset (plotted in Fig. \ref{fig:iris-scatter}) has 4 variables, representing the dimensions of particular features (the sepal and petal lengths and widths) of multiple Iris flower specimens, and includes a total of three iris species/classes (Setosa, Virginica \& Versicolor). The dataset is made up of 150 different flower specimens (50 of each species) with only one of the three species linearly separable from the others by the provided variables. Using the 4 variables, namely Sepal Length (SL), Sepal Width (SW), Petal Length (PL) and Petal Width (PW), as inputs for the spiking RC system, we can effectively create an SNN architecture (depicted in Fig. \ref{fig:reservoir-architecture}) capable of classifying the 150 Iris flowers into their respective 3 species.
As in more traditional laser-based TDRs for photonic RC systems, we make use of time division multiplexing (TDM) to create the spiking (virtual) nodes in the proposed VCSEL-based spiking RC system. Sampling the temporal spiking output of the VCSEL at time intervals of $\theta$, provides the spiking or non-spiking output state of each  node in the SNN. These  nodes are interconnected by the system’s neural-like dynamics (e.g. leaky integrate-and-fire behaviour, refractory period), which occur on timescales around 1\,ns \cite{Hejda2020}. Hence, using a  spiking node duration $\theta$ lower than 1\,ns allows neighbouring nodes to influence one another, creating fixed network connections and therefore forming an SNN architecture with a single VCSEL system. Additionally, unlike many previous realizations of laser-based TDRs, in our SNN architecture, in which we strategically chose not to configure recurrent/feedback connections, the number of  spiking nodes ($N_v$) is flexible, and can be defined and modified at will. Hence, in our proposed system, the number of nodes can be tuned by simply modifying the temporal length and characteristics of the input data waveforms injected into the VCSEL-based photonic spiking RC system. This simple, yet powerful approach allows us to scale the number of  spiking nodes in the photonic SNN as required for different tasks and/or desired performance requirements, while maintaining a highly hardware-friendly single-VCSEL system.

\begin{figure}
    \centering
    \includegraphics[width=3.2in]{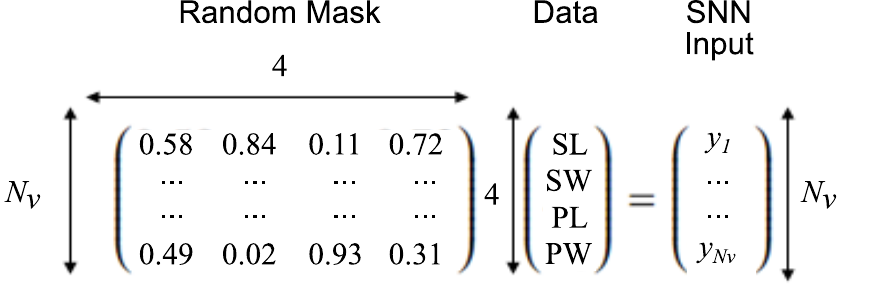}
    \caption{The preparation of the SNN input. The input value for each virtual spiking node is calculated by multiplying the vector of each data point (sepal length (SL), sepal width (SW), petal length (PL) and petal width (PW)) by a random matrix (mask) of length $N_v$.}
    \label{fig:Inputmath}
\end{figure}

\begin{figure}
    \centering
    \includegraphics[width=3.4in]{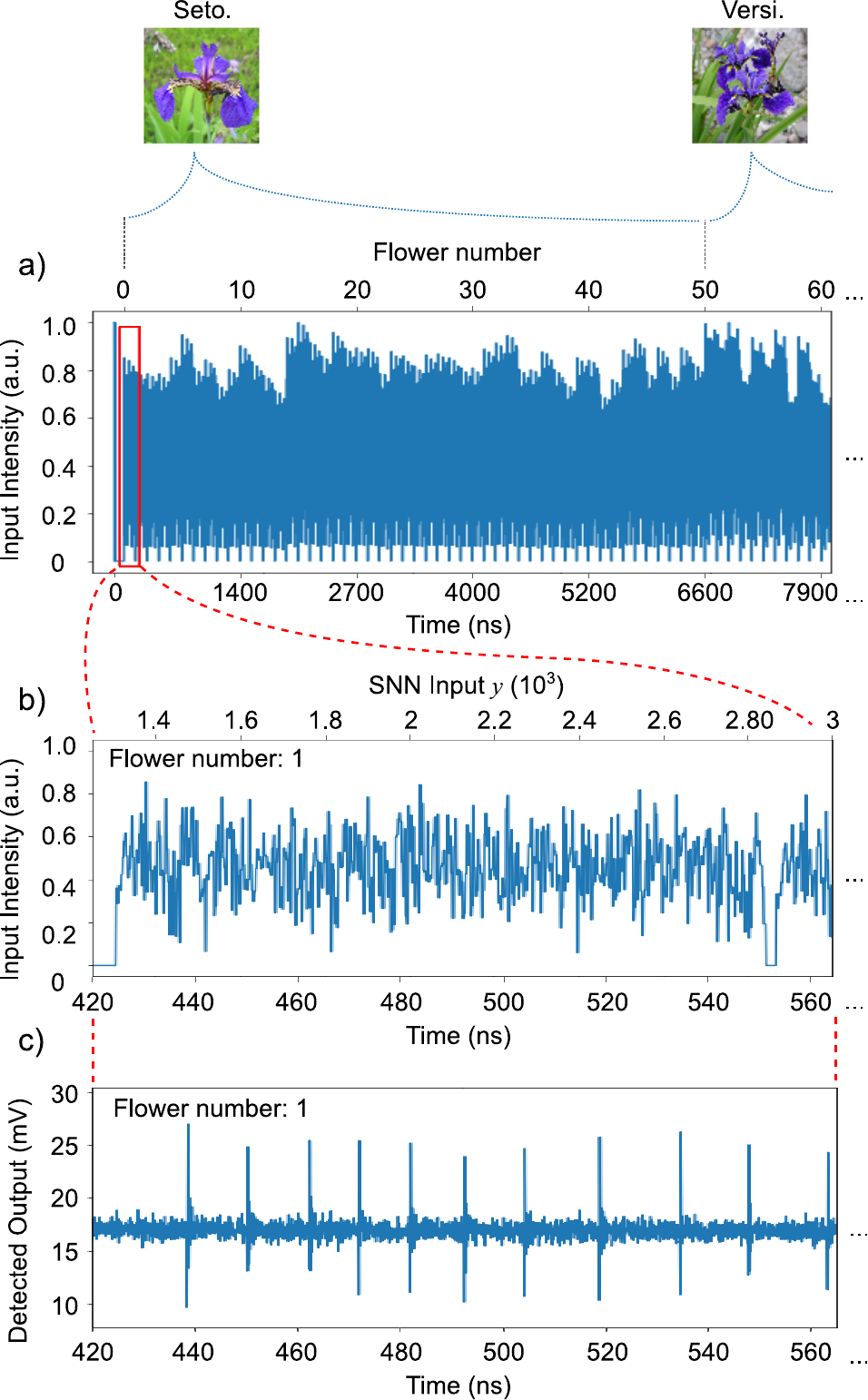}
    \caption{ (a) Section of time series containing the prepared SNN input of all 150 masked data points. The input data points are concatenated by species and separated by 2\,ns. Input shown is for 512 virtual nodes. b) Highlighted red section containing the SNN input of flower number 1. SNN input has been resampled for AWG generation (12\,GSa/s - 3 samples per $\theta$). c) The recorded output timeseries of the SNN, showcasing the spiking pattern obtained for flower number 1.}
    \label{fig:wf-sample}
\end{figure}

Figure \ref{fig:Inputmath} shows how the data is prepared for injection into the VCSEL-based spiking RC system. First a random input weight matrix (mask) is created with 4 columns and as many rows as spiking  nodes ($N_v$) in the SNN. This matrix multiplies a column vector whose components are the 4 Iris flower data point variables (SL, SW, PL, PW), giving a final vector (with $N_v$ components). This vector configures the inputs that are generated by the AWG, encoded in the TL’s light output by the MZ modulator, and optically-injected into the photonic VCSEL-based spiking RC system. Each data point (flower input) is applied the same input weight mask and all 150 flower data points are concatenated in series (as shown in Fig. \ref{fig:wf-sample}\,(a)), with all Setosa Iris flower species directly followed by the Versicolor and Virginica species. Figure \ref{fig:wf-sample}\,(b) shows an expanded section of the input waveform in Fig. \ref{fig:wf-sample}\,(a)), which represents a single input data point (flower number 1)  for the photonic spiking RC system. Time-periods of no input modulation (lasting 2\,ns) are added between each data point (between different flowers) to allow the VCSEL to reach equilibrium between consecutive data inputs (consecutive flowers). These can be seen as the minima values in the input waveform of Fig. \ref{fig:wf-sample}\,(a)), and more clearly after the input data for flower 1 (at approx. SNN Input $y =$ 2.85x$10^3$, time $=$ 550\,ns).

The optical output of the VCSEL, representing the output states of all the spiking nodes in the photonic SNN, is recorded with a real-time oscilloscope (OSC). Figure \ref{fig:wf-sample}\,(c) shows the measured optical output from the VCSEL for the same time segment of input data (flower number 1) included in Fig. \ref{fig:wf-sample}\,(b). The output from the VCSEL-based photonic spiking RC system takes the form of a fast temporal optical spiking pattern (see Fig. \ref{fig:wf-sample}\,(c)), showing multiple (sub-ns long) excitable neuron-like spikes occurring at the output of the different (virtual) spiking nodes of the proposed SNN architecture. To detect which nodes have fired spiking responses the recorded output is cut into bins of $\theta$ (such that there are $N_v$ bins). If the peak power in a bin exceeds the chosen spike threshold, that bin is taken to contain a spike. This process creates a vector of binary (spike \& no-spike) values that are used to produce, using linear-least-squares fitting, the weight matrix of the output layer of the photonic spiking RC system.

\subsection{Experimental Performance of the Photonic Spiking Reservoir Computer}
\label{Sect:3C}

The VCSEL-based SNN implementing the photonic spiking RC system, was configured to run with a spiking virtual node duration of $\theta$ = 250\,ps. Two different configurations with distinct numbers of virtual nodes, namely $N_v$=512 and $N_v$=1024, were tested. In each of these two cases the same procedure was followed at all times, with the only difference being the total duration of each input data point, namely 128 ns/point (for the 512 nodes SNN) and 256 ns/point (for the 1024 nodes SNN). This permitted not only to demonstrate experimentally the successful operation of the photonic spiking RC system of this work, but also to evaluate the system’s performance for different numbers of nodes in the generated SNN.

When testing the system with 512 spiking nodes, the VCSEL was driven with a bias current of 4.02\,mA (approx. 2.9 times the device’s lasing threshold current) and at a constant temperature of 293\,K. The optical injection from the TL was made with 181\,$\mu$W average optical power and an initial frequency detuning of -1.1\,GHz between the externally-injected optical signal and the resonance of the VCSEL’s subsidiary polarization mode. For the system test with 1024 spiking nodes, the driving bias current applied to the VCSEL was equal now to 3.92\,mA (approx. 2.8 times the lasing threshold current), and the optical injection was performed with a 159.4\,$\mu$W average optical input power at a frequency detuning of -3\,GHz between the resonant frequency of the TL’s light and the VCSEL’s subsidiary attenuated mode.

\begin{figure}
    \centering
    \includegraphics[width=3.45in]{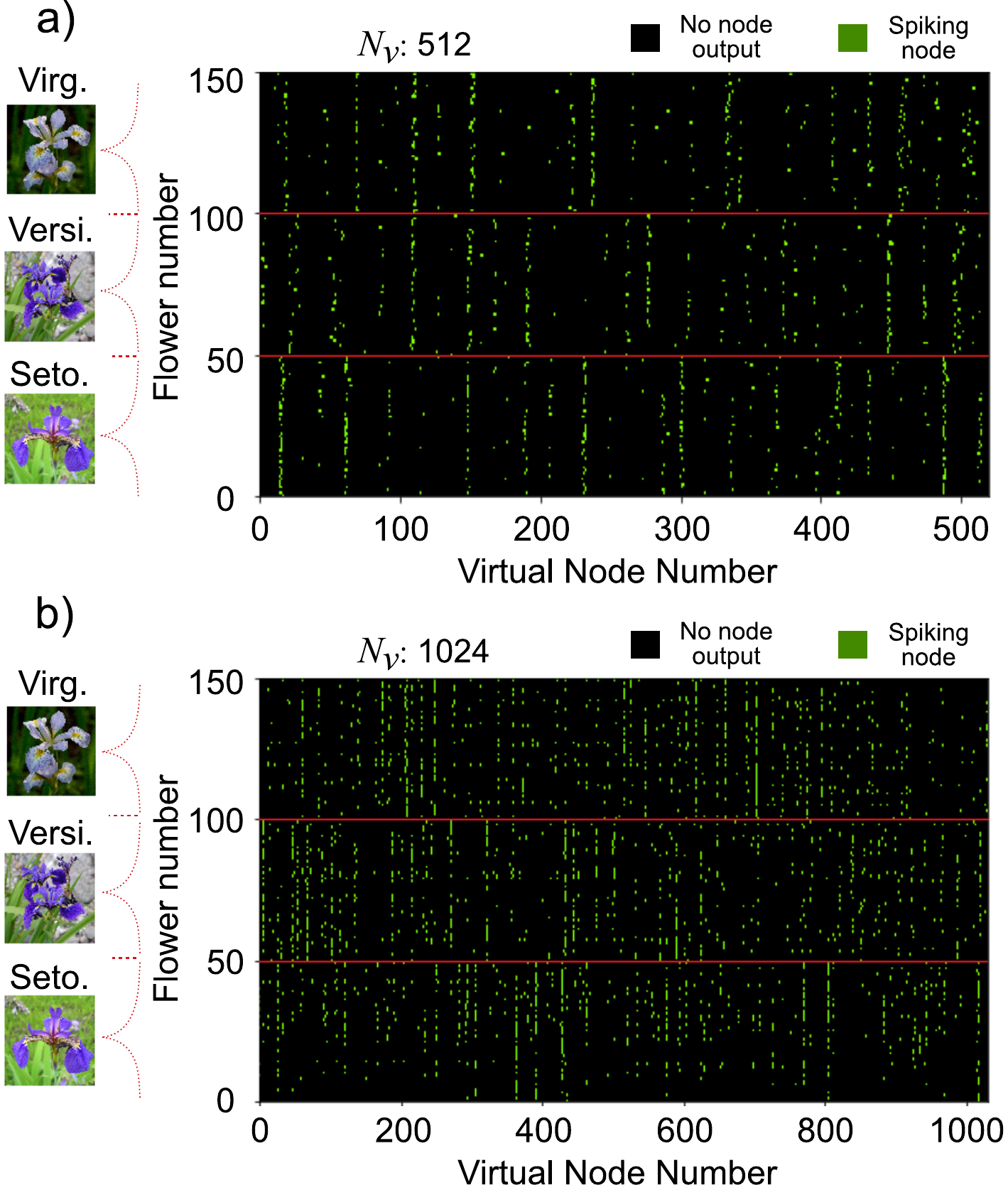}
    \caption{Spike patterns produced by the SNN for all 150 flower data points (50 of each species). The system was tested using (a) 512 and (b) 1024 virtual nodes, with each input flower producing a different spiking pattern. Green represents the presence of a spike within the virtual node.}
    \label{fig:spike-mesh}
\end{figure}

Figure \ref{fig:spike-mesh} plots the experimentally measured temporal maps of all the optical spike patterns produced at the output of the VCSEL-based SNN in response to all 150 data points in the Fisher’s Iris flower dataset. Temporal maps are plotted for both tests of 512 (Fig. \ref{fig:spike-mesh}\,(a)) and 1024 (Fig. \ref{fig:spike-mesh}\,(b)) spiking nodes. In the temporal maps of Fig. \ref{fig:spike-mesh}, the nodes that displayed (did not display) a spike firing event are plotted in green (black). The temporal maps show that for both cases investigated each data point (each input flower specimen) produces a characteristic response, with distinct spiking patterns shared across flowers of the same species (class).

By interpreting the output state of each spiking (virtual) node in the photonic SNN, the weights of the output layer were trained using a linear-least-squares fit. The data used for training was arranged into two matrices: a first matrix in which the rows were the spiking patterns produced by each data point, and a second matrix with three columns to label the data (i.e. a value of ‘1’ in  the first, second or third column denotes the data point as classes ‘1’, ‘2’, and ‘3’, respectively). The output layer weight matrix $W$ is therefore sought in such a way that solves (with the least square error) the equation:

\begin{equation}
    S*W=L,
\end{equation}

Where $L$ is the label matrix and $S$ corresponds to the matrix generated from the spike patterns obtained at the output of the VCSEL SNN. In $S$ the (virtual) nodes with and without elicited optical spikes are represented by binary ‘1’ and ‘0’ values, respectively. $W$ is the weight matrix for the output layer of the spiking RC system (formed by the VCSEL SNN). This can be calculated directly from the product of the Moore-Penrose inverse of $S$ and $L$, giving a matrix with $N_v$ rows and 3 columns ($W=S^{+}*L$). Subsequently, by multiplying the output weight matrix $W$ by the spiking pattern produced by the SNN in response to a new input data point, we produce a triplet containing the probabilities that the data point belongs to each of the three Iris flower classes.

\begin{figure}
    \centering
    \includegraphics[width=2.6in]{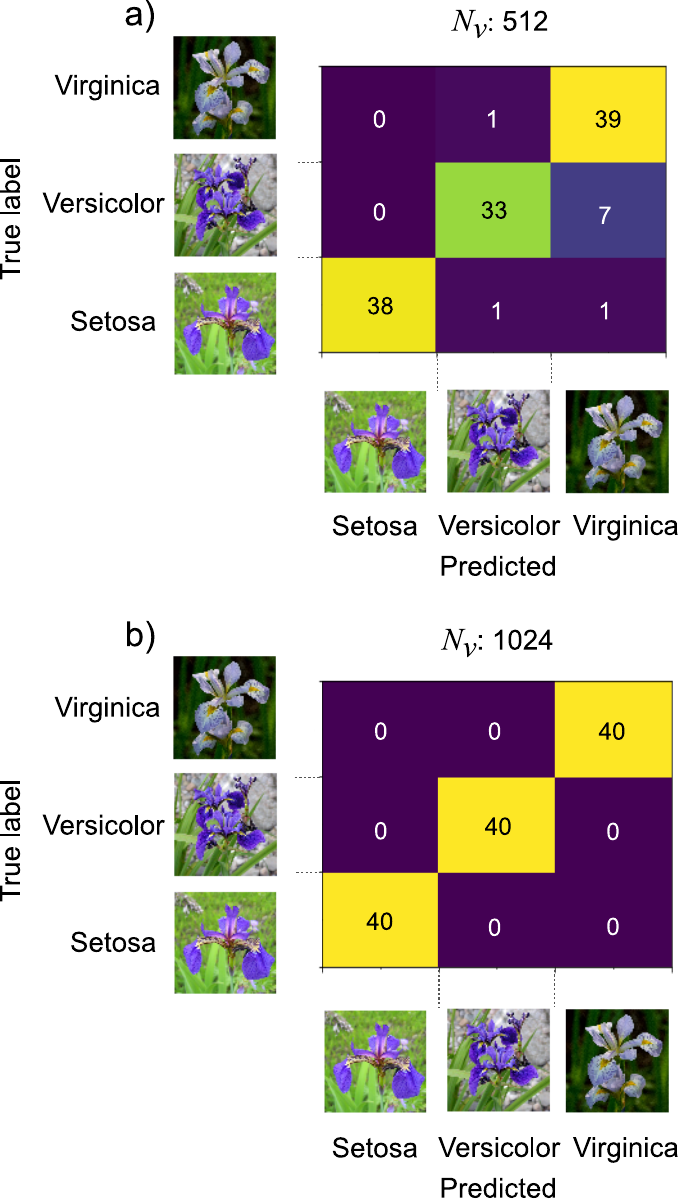}
    \caption{Confusion matrix for the Iris flower classification task using a) 512, and b) 1024 spiking (virtual) nodes in the VCSEL-based SNN implementing the photonic spiking RC system. A training set of 10 data points was used for each class.}
    \label{fig:conf-512}
\end{figure}

Training was done using an equal random selection of data points from each of the three (Iris flower) classes, with the remaining points used to test the accuracy of the VCSEL SNN (photonic spiking reservoir computer). Figure \ref{fig:conf-512} provides confusion matrices obtained from the VCSEL SNN (with 512 or 1024 spiking nodes) when performing the Iris flower classification task. The confusion matrices in Fig. \ref{fig:conf-512} were obtained for a training set size of 30 (10 of each flower class) and a test set size of 120 (40 of each flower). Figure \ref{fig:conf-512} shows the VCSEL-based photonic spiking reservoir's very good performance of the Iris flower classification task. Specifically, Fig. \ref{fig:conf-512}\,(a) shows that the SNN correctly classifies almost all of the flower data points yielding an average classification accuracy (correct classifications divided by the number of test points) of 0.917. Notably, Fig. \ref{fig:conf-512}\,(b) reveals that once the number of nodes in the photonic SNN is increased to 1024, the spiking RC system correctly classifies all 120 tested data points, reaching full accuracy. 

\begin{figure}
    \centering
    \includegraphics[width=3.2in]{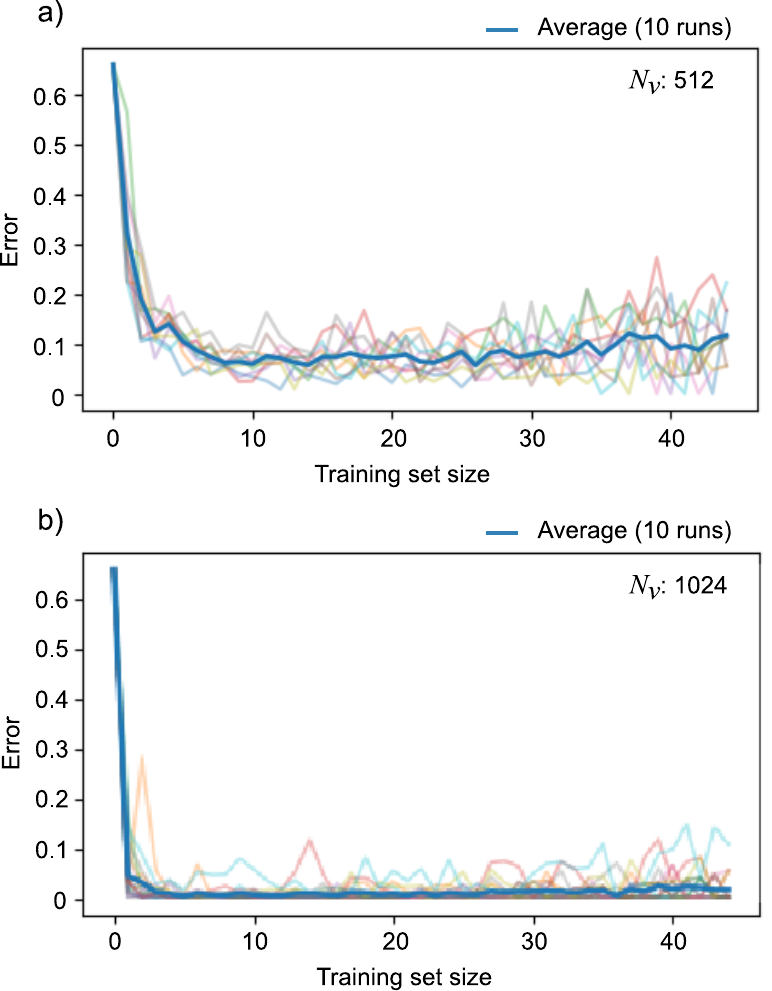}
    \caption{The classification error is plotted for various training set sizes. The error for runs of both (a) 512 and (b) 1024 spiking (virtual) nodes are shown. Faint lines represent several experimental runs of the VCSEL-based photonic SNN, the dark blue solid line represents the average error.}
    \label{fig:training-error}
\end{figure}

Further, we investigate the performance of the VCSEL-based SNN when different training set sizes are applied to both configurations of 512 and 1024 virtual spiking nodes. Here we determine the ideal training dataset size that will allow for the performance optimization of the nonlinear classification task. The curves plotted in Fig. \ref{fig:training-error} provide the system’s classification error obtained across 10 different experimental runs for both cases of 512 and 1024 spiking (virtual) nodes. The calculated average error in each of the two cases is given by the dark blue solid line (see Fig. \ref{fig:training-error}). For both cases, Fig. \ref{fig:training-error} shows that as expected, at first for very small training dataset sizes (up to 3 points) the system’s error is very high. However, remarkably, the system’s performance improves drastically in both cases for very small training set sizes (from values as low as $>$5 training points), reaching classification error minima at around training set sizes of 10. The error remains very low for increasing training dataset sizes, increasing slightly for very large training dataset sizes ($>$25-30 training data points). This increase in error can be explained from the reduced size of the test dataset (as the training dataset size increases). Here, just one or two classification errors will have a larger effect on the total error of the system. For both cases investigated, very high performance is obtained, with classification accuracies remaining consistently over 90\% and 97\% when the VCSEL SNN is configured with 512 or 1024 spiking nodes.
Therefore, the successful operation of a complex nonlinear classification task is demonstrated by the photonic spiking RC system of this work. Notably, very high performance ($>$97\% classification accuracy) is obtained using a system with highly-reduced hardware complexity (which uses a single off-the-shelf, telecom VCSEL), low energy requirements for operation ($\sim$100s of $\mu$Ws optical input power signals and a few mAs of applied bias current) and ultrafast speed operation (250\,ps spiking node separations and approx. 100\,ps-long optical spikes for computation). Moreover, the proposed VCSEL-based SNN architecture permits to modify at will the number of coupled spiking virtual nodes, configuring the photonic spiking RC system depending on the complexity of the task to be performed and the desired levels of performance. The performance of our proposed single-VCSEL SNN compares positively to recently reported traditional (continuous) photonic RC systems \cite{Borghi:2021} when attempting the same benchmark (Iris flower) classification task in terms of accuracy, power requirements and operational speed. Additionally, our VCSEL-based SNN, thanks to its unique operation with ultrafast (approx 100\,ps-long) spikes for computation, removes any requirement for smoothing and down-sampling \cite{Borghi:2021} when obtaining the readout of virtual nodes.

\section{Conclusion}

In summary, we report and experimentally demonstrate for the first time a photonic Spiking Neural Network (SNN) built with just one VCSEL. Our technique, merging in the same platform the excitable spiking dynamics of artificial photonic (VCSEL-based) neurons with the RC/ELM paradigm, describes a radically new class of photonic RC system using neural-like optical spikes to compute (thus yielding a truly neuromorphic photonic processor). In our approach, the optical spiking output of the VCSEL RC system, is time-multiplexed into multiple  temporal spiking virtual nodes. This creates a photonic SNN with the spiking nodes interconnected through the non-linear integrate-and-fire and refractory dynamics of the excitable neuromorphic (spiking) responses elicited by the VCSEL. We reveal the successful operation of this first photonic spiking RC system on a complex nonlinear classification task. Notably, very high accuracy was achieved ($>$97\%) in the so-called Iris flower classification task, thus showcasing its powerful computational performance. It is also important to mention that the spiking photonic RC system of this work is built using a single commercially-available telecom VCSEL and off-the shelf fibre-optic components; hence benefiting from an extremely simple, inexpensive and hardware-friendly implementation. Additionally, the system of this work also offers other important inherent attributes, including ultrafast (GHz rates, 250\,ps/node) and low-power operation ($\sim$100s of $\mu$Ws optical input power, few mAs of VCSEL bias current). It is also worth mentioning that the characteristics of the photonic SNN implementing the (VCSEL-based) photonic spiking RC system can be modified at will. In the proof-of-concept demonstration of this work, we showcase the possibility to fully control the number of interconnected spiking nodes in the photonic SNN. In our demonstration, we report results for 512 and 1024 spiking nodes (at 250\,ps/node), but SNNs with either a smaller or a larger number of spiking nodes are possible, depending on the complexity of the task to be processed and the desired performance requirements (e.g. accuracy levels desired, total operation speed). It is also important to note that whilst we report and experimentally demonstrate the first photonic spiking RC system using a VCSEL as the core nonlinear (spiking) element, the architecture described could be also transferred to other excitable spiking systems (photonic or otherwise) to deliver a new generation of spiking RC systems based upon different technologies. These results therefore open exciting new avenues towards ultrafast, low-power photonic SNNs yielding powerful processing hardware-systems for future light-enabled neuromorphic computing and AI platforms.

\bibliographystyle{IEEEtran}
\bibliography{manuscript_1}


\begin{IEEEbiographynophoto}{Dafydd Owen-Newns}
recieved a MPhys Physics from the University of Exeter in 2021, and is currently working towards a Ph.D. at the Institute of Photonics, University of Strathclyde, Glasgow, focusing on neuromorphic computing using lasers and optoelectronic devices.
\end{IEEEbiographynophoto}

\begin{IEEEbiographynophoto}{Joshua Robertson}
received his MPhys degree in Physics with specialization in Photonics from the University of Strathclyde, Glasgow in 2017. He is currently finalizing his Ph.D. degree in Physics at the Institute of Photonics (IoP), University of Strathclyde, Glasgow. His studies investigate and report on neuromorphic photonic systems with lasers, with focus on the experimental spiking devices for photonic-based computing.
\end{IEEEbiographynophoto}

\begin{IEEEbiographynophoto}{Mat\v{e}j Hejda}
received the Ing. degree (MSc. equivalent) in Nanotechnology from Technical University of Liberec, Czech Republic in 2019. Currently, he's pursuing a Ph.D. degree in Physics in the Neuromorphic Photonics group at the Institute of Photonics, University of Strathclyde, Glasgow, U.K. His main focus is on investigation of spiking in photonic and optoelectronic devices towards applications in light-powered neuromorphic computing.
\end{IEEEbiographynophoto}

\begin{IEEEbiographynophoto}{Antonio Hurtado}
is a Reader and Turing Artificial Intelligence (AI) Fellow at the University of Strathclyde’s Institute of Photonics (IoP). He received the Ph.D. degree from the Universidad Politécnica de Madrid (UPM), Madrid, Spain in December 2006. He has over 15 years’ international research experience in photonics in the UK (Universities of Essex and Strathclyde), USA (University of New Mexico), and Spain (UPM). He received two Marie Curie Fellowships by the European Commission: Projects ISLAS (2009–2011) and NINFA (2011–2014). In 2014 he was awarded Chancellor’s Fellowship by the University of Strathclyde following which he was appointed as a Lecturer in the Strathclyde’s Institute of Photonics (Physics Dept.). He was promoted to Senior Lecturer (2018) and Reader (2021). His current research interests are but not limited to neuromorphic photonics, laser nonlinear dynamics, nanolaser systems and hybrid nanofabrication. In 2020 he was awarded a Turing AI Acceleration Fellowship by the UK Research and Innovation (UKRI) office and the UK Government Business, Energy and Industrial Strategy (BEIS) Department to develop a 5-year research programme on Photonics for Ultrafast AI.
\end{IEEEbiographynophoto}





\end{document}